\documentclass[aps,eqsecnum,preprint,preprintnumbers,12pt]{revtex4}
\usepackage{epic}

\begin{document}

\preprint{SNUTP 02-029}
\title{\hspace{1cm}\\\hspace{1cm}\\\hspace{1cm}\\Semiclassical Theory for Two-anyon System}
\author{Jin Hur} \email{hurjin2@snu.ac.kr}
\author{Choonkyu Lee} \email{cklee@phya.snu.ac.kr}
\affiliation{Department of Physics and Center for Theoretical
Physics\\ Seoul National University, Seoul 151-742,
Korea\\\hspace{1cm}\\\hspace{1cm}\\}

\begin{abstract}
The semiclassical quantization conditions for all partial waves
are derived for bound states of two interacting anyons in the
presence of a uniform background magnetic field. Singular
Aharonov-Bohm-type interactions between the anyons are dealt with
by the modified WKB method of Friedrich and Trost. For s-wave
bound state problems in which the choice of the boundary condition
at short distance gives rise to an additional ambiguity, a
suitable generalization of the latter method is required to
develop a consistent WKB approach. We here show how the related
self-adjoint extension parameter affects the semiclassical
quantization condition for energy levels. For some simple cases
admitting exact answers, we verify that our semiclassical formulas
in fact provide highly accurate results over a broad quantum
number range.
\end{abstract}
\maketitle

\section{introduction}

In two-dimensional space where the rotation group (SO(2)) is
abelian, particles obeying fractional statistics, i.e., anyons,
can exist \cite{leinaas, wilczek, khare, lerda}. They can be given
a concrete mathematical representation in flux-charge composites,
since `charged' bosonic (or fermionic) particles with non-zero
magnetic flux attached behave like anyons thanks to the
Aharonov-Bohm type interference \cite{aharonov}. Among the
physical realizations of anyonic objects, the quasiparticle or
quasihole in the Laughlin state of the fractional quantum Hall
system corresponds to a particularly interesting case
\cite{arovas}. In this regard, a challenging problem is to settle
from the first principles what sort of quantum states would be
allowed for anyon matter with some realistic interactions (and
possibly a uniform background magnetic field).

Choosing {\it bosonic} description, quantum dynamics of a system
of anyons will be governed by the Hamiltonian of the form
\cite{khare, lerda}
\begin{equation}
H = \sum_{n} \frac{1}{2m} \left[ \vec{p}_{n} - \alpha \vec{a}_{n}
- e \vec{A}^{\rm ex}(\vec{r}_{n}) \right]^{2} + \frac{1}{2}
\sum_{n,m(\neq n)} V(|\vec{r}_{n}-\vec{r}_{m}|),
\end{equation}
where $\vec{r}_{n} = ( x_{n}^{1} ,x_{n}^{2} ), \; \vec{A}^{\rm
ex}(\vec{r}) = \displaystyle \frac{B}{2} ( -x^{2}, x^{1} )$ with
$B>0$ is the vector potential appropriate to a uniform background
magnetic field, and
\begin{equation} \label{vpot}
a^{i}_{n} = \epsilon^{ij} \sum_{m(\neq n)}
\frac{x^{j}_{n}-x^{j}_{m}}{|\vec{r}_{n}-\vec{r}_{m}|^{2}}.
\end{equation}
The parameter $\alpha$ characterizes the type of anyons and
without loss of generality $\alpha$ may be restricted to the
interval (-1, 1]. For $\alpha=0$ ($\alpha=1$), we have physical
bosons (fermions). This defines a Galilean-invariant system if the
background magnetic field, included for generality, is turned off
(i.e., for $B=0$). In the presence of a generic two-body
interaction potential $V(|\vec{r}_{n}-\vec{r}_{m}|)$ together with
the Aharonov-Bohm interactions represented by the point vortex
potentials (\ref{vpot}), the above Hamiltonian leads to a
nontrivial energy eigenvalue equation even for the two-body case
--- the starting point of a systematic n-body study. [See Refs.
\cite{khare, lerda, johnson, wu, dunne} where some related studies
are made, usually without a two-body potential term]. For the
two-anyon system the time-independent Schr\"{o}dinger equation,
after separating the center-of-mass dynamics (just the problem of
a particle with mass $M=2m$ and charge $2e$ moving in the
background magnetic field), reduces to the following equation
which has only the relative position $\vec{r} = \vec{r}_{2} -
\vec{r}_{1} \equiv (x, y)$ as independent variables:
\begin{equation} \label{eom}
\left\{ -\frac{1}{2\mu} \left[ \vec{\nabla}_{r} - i \alpha
\vec{a}(\vec{r}) - \frac{ie}{2} \vec{A}^{\rm ex}(\vec{r})
\right]^{2} + V(r) \right\} \psi(\vec{r}) = E \psi(\vec{r}),
\end{equation}
where $\mu = \displaystyle \frac{1}{2} m$ and $a^{i}(\vec{r}) =
\epsilon^{ij} \displaystyle \frac{x^{j}}{r^{2}}$. [We set $\hbar =
1$ in this paper]. The relative dynamics is effectively that of a
single particle moving in the presence of a uniform magnetic
field, a point vortex at the origin, and a certain radial
potential $V(r)$.

With a nontrivial potential $V(r)$ an exact analysis of
(\ref{eom}) is usually not possible and hence suitable
approximation methods may be sought. In this paper, we shall study
quantum bound states of two anyons through the semiclassical or
WKB analysis of the radial Schr\"{o}dinger equations for partial
wave amplitudes $\psi_{l}(r) \; (l=0, \pm 2, \pm 4, \cdots)$ which
are derived from (\ref{eom}). To account for the effects due to
the singular point vortex potential (as well as the centrifugal
potential term) within the WKB method, the conventional Langer
modification \cite{langer} of the potential is not adequate; but,
the modified method of Friedrich and Trost \cite{friedrich} can be
used in an effective way. Based on the latter method, we have
obtained the semiclassical quantization condition (containing a
nonintegral Maslov index \cite{gutzwiller})
\begin{equation} \label{qc}
\int_{r_{1}}^{r_{2}} dr \sqrt{ 2\mu ( E - V_{\it eff}(r) ) } =
\left( n + \frac{1}{2} + \frac{1}{2} |l-\alpha| - \frac{1}{2}
\sqrt{\eta} \right) \pi, \;\;\;\;\; ( n = 0, 1, 2, \cdots )
\end{equation}
where $\eta \equiv (l-\alpha)^{2} - \displaystyle \frac{1}{4}$
(assumed to be positive here), $V_{\it eff}(r)$ is the effective
one-dimensional potential
\begin{equation} \label{effv}
V_{\it eff}(r) = V(r) - \frac{eB}{4\mu} (l-\alpha) +
\frac{e^{2}}{32\mu} B^{2} r^{2} + \frac{\eta}{2\mu r^{2}},
\end{equation}
and $r_{1}$ and $r_{2}$ refer to the related classical turning
points. [The condition (\ref{qc}) may be used when the WKB wave
function in the region $r>r_{2}$ contains a decreasing exponential
only]. Efficacy of this formula becomes evident once one sees how
its predictions compare against the exact values available for
some simple cases.

Actually, for the case of the s-wave amplitude $\psi_{l=0}(r)$,
there exists an additional complication involving the choice of
boundary condition at $r=0$ (i.e., at the point of two-particle
coincidence). We know from the theory of self-adjoint extension
that there exist a one-parameter family of acceptable boundary
conditions \cite{albeverio}, including the so-called hard-core
boundary condition \cite{aharonov} as a special case. [Note that
there is no {\it a priori} reason to choose specifically the
hard-core boundary condition --- a real system under study should
determine the relevant boundary condition eventually]. In
accordance with this theory, one can represent the s-wave
amplitude $\psi_{l=0}(r)$ for small $r$ by the form
\begin{equation} \label{other}
\psi_{l=0}(r) \propto \left\{ J_{|\alpha|}(kr) + \tan \theta
\left( \frac{k}{\rho} \right)^{2|\alpha|} J_{-|\alpha|}(kr)
\right\},
\end{equation}
where $k \equiv \sqrt{ 2\mu ( E - V(0) - \frac{eB}{4\mu} \alpha )
}$, $\theta$ (the self-adjoint extension parameter) is a real
dimensionless number, and $\rho$ a reference scale introduced for
convenience. The hard-core boundary condition is associated with a
special choice, $\theta=0$. Needless to say, this
boundary-condition-dependent effect should be taken into account
in the WKB analysis of the s-wave amplitude. The Langer
modification method is simply not applicable here; but, the method
of Friedrich and Trost has a natural generalization for this
problem. The resulting s-wave semiclassical quantization condition
we have obtained for $\displaystyle \frac{1}{2} \leq |\alpha| \leq
1$ has the form
\begin{equation}
\int_{r_{1}}^{r_{2}} dr \sqrt{ 2\mu ( E - V_{\it eff}(r) ) } =
\left( n + \frac{1}{2} + \frac{1}{2} |\alpha| - \frac{1}{2} \sqrt{
\alpha^{2} - \textstyle \frac{1}{4} } \right) \pi + \Theta(E),
\;\;\;\;\; ( n = 0, 1, 2, \cdots )
\end{equation}
where $\Theta(E)$, a function of energy $E$ (through $k = \sqrt{
2\mu ( E - V(0) - \frac{eB}{4\mu} \alpha ) }$ ), is related to the
above self-adjoint extension parameter by
\begin{equation} \label{cthetas}
\tan\Theta = - \frac{ \sin\pi|\alpha| \tan\theta \;
(k/\rho)^{2|\alpha|} }{ 1 + \cos\pi|\alpha| \tan\theta \;
(k/\rho)^{2|\alpha|} }.
\end{equation}
For $|\alpha| < \displaystyle \frac{1}{2}$, we need some extra
consideration and this case is covered by our another formula,
given in (\ref{s-qc}).

It might be of some interest to look at our problem also from the
viewpoint of field theory. In the field-theoretic description, one
can describe (nonrelativistic) anyons by using bosonic
Schr\"{o}dinger fields $\psi(\vec{r},t), \;
\psi^{\dag}(\vec{r},t)$ coupled to an abelian Chern-Simons gauge
field $a_{\mu}(\vec{r} ,t) \; (\mu = 0, 1, 2)$ \cite{hagen}. For
the above system specifically, one may consider the Lagrangian
density
\begin{eqnarray} \label{ld}
{\cal L} = \frac{\kappa}{2} \epsilon^{\mu\nu\lambda} a_{\mu}
\partial_{\nu} a_{\lambda} + \psi^{\dag} (i D_{t} + \frac{1}{2m}
\vec{D}^{2}) \psi - \frac{1}{2} \lambda_{B}
\psi^{\dag} \psi^{\dag} \psi \psi \nonumber \\
- \frac{1}{2} \int d^{2}\vec{r}\;' \psi^{\dag}(\vec{r},t)
\psi^{\dag}(\vec{r}\;',t) V(|\vec{r}-\vec{r}\;'|)
\psi(\vec{r}\;',t) \psi(\vec{r},t),
\end{eqnarray}
where $D_{t} = \partial_{t} + i q a_{0}, \; \vec{D} = \vec{\nabla}
- i q \vec{a} - i e \vec{A}^{\rm ex}$, and the anyon parameter
$\alpha$ should be identified with $\displaystyle
\frac{q^{2}}{2\pi \kappa}$. Here the contact interaction term,
$\displaystyle -\frac{1}{2} \lambda_{B} \psi^{\dag} \psi^{\dag}
\psi \psi$, which becomes necessary to ensure the
renormalizability of the theory, is responsible for the s-wave
boundary condition ambiguity mentioned above \cite{bergman}; for
the precise correspondence between the method of self-adjoint
extension in quantum mechanics and the
regularization/renormailization procedure in field theory, see
Refs. \cite{bergman, kim, kim2}. Thus, with a suitable
transcription made from the self-adjoint extension parameter to
the renormalized contact coupling $\lambda_{R}$ (as considered,
for instance, in Ref. \cite{kim2}), our WKB analysis provides
appropriate results for the two-particle bound states of this
nonrelativistic field theory system also. But, in this work, such
field theoretic language will not be used.

This paper is organized as follows. In Sec.2 we use the modified
WKB method to find approximate non-s-wave bound states of the
Schr\"{o}dinger equation (\ref{eom}) and especially derive the
semiclassical energy level formula (\ref{qc}). Then, for some
simple cases (e.g., for a circular billiard), we check our
WKB-based predictions against the exact results. In Sec.3 the case
of s-wave bound states is studied within the framework of the
modified WKB method, with special attention given to the
dependence on the self-adjoint extension parameter (or contact
coupling). Section 4 contains a summary and discussions of our
work.

\section{non-s-wave semiclassical bound states}

The relative dynamics of the two anyon system is governed by the
Schr\"{o}dinger equation (\ref{eom}). This equation is analyzed
most conveniently in polar coordinates, as the vector potentials
in our problem are equal to
\begin{equation}
\vec{a}(\vec{r}) = \frac{1}{r} \hat{\theta} \;\;\;\;\; ,
\;\;\;\;\; \vec{A}^{\rm ex}(\vec{r}) = \frac{B}{2} r \hat{\theta}.
\end{equation}
Now, writing $\psi(\vec{r}) = \psi_{l}(r) e^{il\theta}$ (with $l$
restricted to even integer values in our bosonic description of
anyons) in (\ref{eom}), one obtains the following radial equation
for the partial wave amplitude $\psi_{l}(r)$:
\begin{equation} \label{reom}
\left\{ -\frac{1}{2\mu} \left[ \frac{d^{2}}{dr^{2}} + \frac{1}{r}
\frac{d}{dr} - \frac{1}{r^{2}} (l-\alpha)^{2} \right] +
\tilde{V}(r) \right\} \psi_{l}(r) = E \psi_{l}(r),
\end{equation}
where $\tilde{V}(r) \equiv V(r) - \displaystyle \frac{eB}{4\mu}
(l-\alpha) + \frac{e^{2}B^{2}}{32\mu} r^{2}$. Introducing the
function $R_{l}(r)$ by
\begin{equation}
\psi_{l}(r) = \displaystyle \frac{1}{\sqrt{r}} R_{l}(r),
\end{equation}
(\ref{reom}) can be further recast into the form
\begin{equation} \label{reom2}
\left\{ -\frac{1}{2\mu} \frac{d^{2}}{dr^{2}} + V_{\it eff}(r)
\right\} R_{l}(r) = E R_{l}(r)
\end{equation}
with the effective one-dimensional potential (given already in
(\ref{effv}))
\begin{equation} \label{veff}
V_{\it eff}(r) = \tilde{V}(r) + \frac{\eta}{2\mu r^{2}},
\;\;\;\;\; (\eta \equiv (l-\alpha)^{2} - \frac{1}{4}).
\end{equation}
The eigenvalue problem (\ref{reom2}) for $l = \pm 2, \pm 4,
\cdots$ (and hence $\eta>0$, assuming $|\alpha| \leq 1$) will be
studied by the semiclassical method in this section. The $l=0$
case is considered separately in the next section.

A naive application of the WKB method with the differential
equation (\ref{reom2}) does not yield very satisfactory results
(especially for relatively small $l$), the reason being that (i)
the radial coordinate $r$ runs from 0 to $\infty$ (instead of
$-\infty$ to $\infty$) and (ii) the potential contains a singular
term at the origin, $\displaystyle \frac{\eta}{2\mu r^{2}}$. In
fact, the naive WKB wave function does not even have the correct
small-$r$ behavior. This defect gets significantly reduced if one
introduces, within the usual WKB approach, the so-called Langer
modification of the potential \cite{langer}, effected through
replacing the potential (\ref{veff}) by
\begin{equation}
V_{\it eff}^{L}(r) = \tilde{V}(r) + \frac{ \eta + \frac{1}{4} }{
2\mu r^{2} }.
\end{equation}
But, to go beyond this simple Langer modification scheme, it
becomes necessary to incorporate the {\it correct phase loss} due
to reflection at a classical turning point in the WKB wave
function corresponding to the classically allowed region. This
leads to the modified WKB method, as described recently by
Friedrich and Trost \cite{friedrich}. In this paper, we shall use
the latter method for our semiclassical discussion and see how the
resulting predictions compare against exact results (and also
those obtained with the Langer modification) for some special
cases.

\begin{figure}[t]
\begin{picture}(230,180)
\put(210,0){$r$}
\put(210,100){$E$}
\put(0,170){$V_{\it eff}(r)$}
\put(5,0){$r_{1}$}
\put(156,0){$r_{2}$}
\put(0,10){\vector(0,1){150}}
\put(0,10){\vector(1,0){225}}
\put(0,110){\line(1,0){225}}
\put(9.5,10){\dashline[10]{3}(0,0)(0,100)}
\put(161,10){\dashline[10]{3}(0,0)(0,100)}
\thicklines \bezier{5000}(5,160)(15,-90)(200,160)
\end{picture}
\caption{The shape of our potential $V_{\it eff}(r)$ with $\eta >
0$. \label{fig}}
\end{figure}
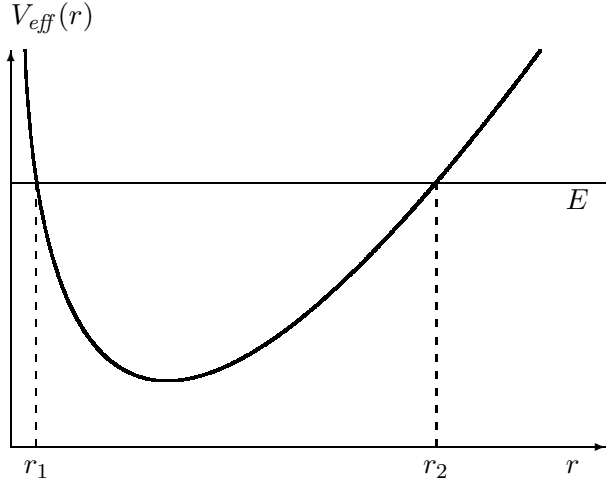

We may here suppose that the potential $V_{\it eff}(r)$, as given
by (\ref{veff}) (with $\eta > 0$), has a typical shape shown in
Fig. \ref{fig}.
Also it may be assumed that the potential $V_{\it eff}(r)$ near
the origin can be approximated by $\tilde{V}(0) + \displaystyle
\frac{\eta}{2\mu r^{2}}$, with $\tilde{V}(0) = V(0) -
\displaystyle \frac{eB}{4\mu} (l-\alpha)$. Then the exact solution
of (\ref{reom2}), which is regular at the origin, should have the
small-$r$ behavior as given by
\begin{equation} \label{ex}
R_{l}^{\rm ex}(r) \propto \sqrt{kr} J_{\nu}(kr),
\end{equation}
where $k = \sqrt{ 2 \mu ( E - \tilde{V}(0) ) }$, and $J_{\nu}$ is
the Bessel function of order $\nu \equiv \sqrt{ \eta + \frac{1}{4}
} = |l-\alpha|$. We are here interested in finding the WKB bound
states in the situation where there are two classical turning
points (see Fig. \ref{fig}), $r_{1}$ and $r_{2}$. The WKB wave
function in the classically allowed region $r_{1} < r < r_{2}$ may
be written as
\begin{equation} \label{WKB}
R_{l}^{\rm WKB}(r) \propto \frac{1}{\sqrt{p(r)}} \cos \left\{
\int_{r_{1}}^{r} dr' p(r') + \phi \right\},
\end{equation}
where $p(r) \equiv \sqrt{ 2 \mu ( E - V_{\it eff}(r) ) } = \sqrt{
k^{2} - \displaystyle \frac{\eta}{r^{2}} - 2 \mu ( \tilde{V}(r) -
\tilde{V}(0) ) }$, and the phase $\phi$ is yet undetermined. Note
that $2\phi$ can be identified as the phase loss of waves due to
reflection by the barrier at $r<r_{1}$. In the conventional WKB
method, one then uses the famous connection formula to combine the
oscillating wave function in (\ref{WKB}) with suitable
monotonically decreasing wave functions in the classically
forbidden regions (i.e., $0<r<r_{1}$ and $r>r_{2}$). But, by the
reasons mentioned above, a different strategy must be adopted in
our case in dealing with the WKB wave function for small values of
$r$. Here, following Ref. \cite{friedrich}, we will simply fix the
phase $\phi$ in (\ref{WKB}) in such a way that the asymptotic
expansion of the expression (\ref{WKB}) for relatively large $kr$
may match that following from (\ref{ex}) (i.e., the correct
expression when $r$ is not too large), say, up to terms of order
$\displaystyle \frac{1}{(kr)^{2}}$. On the other hand, in the
vicinity of the second turning point $r=r_{2}$, the normal WKB
method with the standard connection formula may well be used.

Using (\ref{ex}), the exact solution of the Schr\"{o}dinger
equation near the origin should have the asymptotic expansion
\begin{eqnarray} \label{aex}
R_{l}^{\rm ex}(r) &\sim& \left( 1 - \frac{\eta(\eta-2)}{8(kr)^{2}}
\right) \cos \left[ kr - \left( \nu+\frac{1}{2} \right)
\frac{\pi}{2} \right] \nonumber \\
&& - \frac{\eta}{2kr} \sin \left[ kr - \left( \nu+\frac{1}{2}
\right) \frac{\pi}{2} \right] + O \left( \frac{1}{(kr)^{3}}
\right).
\end{eqnarray}
For the asymptotic expansion of the WKB wave function (\ref{WKB}),
we may use the small-$r$ approximation of $V_{\it eff}(r)$, i.e.,
$V_{\it eff}(r) \simeq \tilde{V}(0) + \displaystyle
\frac{\eta}{2\mu r^{2}}$ in $p(r)$ (and correspondingly the
turning point value $r_{1} \simeq \displaystyle
\frac{\sqrt{\eta}}{k}$), and then the integral $\displaystyle
\int_{r_{1}}^{r} dr' p(r')$ can be performed explicitly. The
result is the following asymptotic behavior:
\begin{eqnarray} \label{aWKB}
R_{l}^{\rm WKB}(r) &\sim& \left( 1 -
\frac{\eta(\eta-2)}{8(kr)^{2}} \right) \cos \left[ kr - c + \phi
\right] \nonumber \\
&& - \frac{\eta}{2kr} \sin \left[ kr - c + \phi \right] + O \left(
\frac{1}{(kr)^{3}} \right),
\end{eqnarray}
where $c = \sqrt{\eta} \displaystyle \frac{\pi}{2}$. Comparing
(\ref{aWKB}) with (\ref{aex}), we see that the two expressions
agree up to terms of order $\displaystyle \frac{1}{(kr)^{2}}$ only
if we choose the phase $\phi$ as
\begin{equation} \label{phi}
\phi = c - \left( \nu + \frac{1}{2} \right) \frac{\pi}{2} = \left(
\sqrt{ \mathstrut \eta } - \sqrt{ \mathstrut \eta + \textstyle
\frac{1}{4} } - \frac{1}{2} \right) \frac{\pi}{2}.
\end{equation}
[In contrast, we note that the standard WKB method with the Langer
modification produces the right argument of sine and cosine in
(\ref{aWKB}), but {\it wrong} coefficients for the terms
proportional to $\displaystyle \frac{1}{kr}$ and $\displaystyle
\frac{1}{(kr)^{2}}$ in (\ref{aWKB})].

The phase $\phi$ in our WKB wave function (\ref{WKB}) has been
determined now. On the other hand, the WKB wave function obtained
by applying the standard connection formula at the second turning
point $r=r_{2}$ has the form
\begin{equation} \label{WKB2}
R_{l}^{\rm WKB}(r) \propto \frac{1}{\sqrt{p(r)}} \cos \left( -
\int_{r}^{r_{2}} dr' p(r') + \frac{\pi}{4} \right).
\end{equation}
The two functions (\ref{WKB}) and (\ref{WKB2}) must of course be
the same. From this follows the quantization condition for energy
levels
\begin{eqnarray} \label{WKBcon}
\int_{r_{1}}^{r_{2}} dr' p(r') &=& n\pi - \phi + \frac{\pi}{4}
\nonumber \\
&=& \left[ n + \frac{1}{2} + \frac{1}{2} \left( \sqrt{ \mathstrut
\eta + \textstyle \frac{1}{4} } - \sqrt{ \mathstrut \eta } \right)
\right] \pi. \;\;\;\;\; ( n = 0, 1, 2, \cdots )
\end{eqnarray}
This is our formula (\ref{qc}). For the sake of comparison, we
note that the WKB quantization condition obtained with the help of
the Langer modification reads
\begin{equation} \label{langercon}
\int_{r_{1}}^{r_{2}} dr \sqrt{ 2\mu \left( E - \tilde{V}(r) -
\frac{\eta+\frac{1}{4}}{2\mu r^{2}} \right) } = \left( n +
\frac{1}{2} \right) \pi. \;\;\;\;\; ( n = 0, 1, 2, \cdots )
\end{equation}

We also remark that, for an effective potential having a
qualitatively different form from that shown in Fig. \ref{fig},
the WKB wave function beyond the point $r=r_{2}$ might not be
given by a decreasing exponential only. In such case the phase
$\displaystyle \frac{\pi}{4}$ in the argument of cosine in
(\ref{WKB2}) should be replaced by an appropriate different value,
and our formula for the semiclassical energy levels need to be
adjusted accordingly.

We will now consider some special cases to see how well the
predictions based on our semiclassical formula (\ref{WKBcon}) (and
(\ref{langercon}) also for comparison) fare against the exact
results. Suppose we have $V(r)=0$, but $B \neq 0$, i.e., the case
of noninteracting anyons in a uniform magnetic field
\cite{johnson, khare}. In this case the solution to the radial
equation (\ref{reom2}), which is regular at the origin, can be
expressed in terms of the confluent hypergeometric function
\cite{landau},
\begin{equation} \label{exsol}
R_{l}(r) = \left( \frac{eB}{4} r^{2} \right)^{ \frac{b}{2} -
\frac{1}{4} } e^{ - eBr^{2}/8 } F(a, b, \frac{eB}{4} r^{2}),
\end{equation}
where $a = \displaystyle \frac{1}{2} [ 1 + |l-\alpha| + (l-\alpha)
] - \frac{2\mu E}{eB}$, and $b = 1 + |l-\alpha|$. The large-$x$
asymptotics of the confluent hypergeometric function $F(a,b,x)$ is
well-known: $F(a,b,x) \sim \displaystyle
\frac{\Gamma(b)}{\Gamma(a)} e^{x} x^{a-b}$ if $a \neq 0, -1, -2,
\cdots$, while $F(a,b,x)$ for $a = 0, -1, -2, \cdots$ reduces to a
polynomial (called the associate Laguerre polynomial). Hence, to
obtain a normalizable solution from (\ref{exsol}), the energy $E$
must be restricted to the values
\begin{equation} \label{exharm}
E = \left( n + \frac{1}{2} + \frac{1}{2} |l-\alpha| - \frac{1}{2}
(l-\alpha) \right) \omega_{c}, \;\;\;\;\; ( n = 0, 1, 2, \cdots )
\end{equation}
where $\omega_{c} = \displaystyle \frac{eB}{2\mu}$ is the
classical cyclotron frequency. These are exact energy levels,
exhibiting very different $l$-dependences for $l>0$ and for $l<0$
\cite{khare, johnson}. On the other hand, within our semiclassical
approach, the energy levels follow immediately once we evaluate
the integral in the left hand side of (\ref{WKBcon}). With
$\tilde{V}(r) = - \displaystyle \frac{eB}{4\mu} (l-\alpha) +
\frac{e^{2}B^{2}}{32\mu} r^{2}$, the given integral is readily
evaluated to yield the condition of the form
\begin{eqnarray}
&& \frac{2\pi\mu}{eB} \left\{ E + \frac{eB}{4\mu} (l - \alpha)
\right\} - \sqrt{\eta} \frac{\pi}{2} \nonumber \\
&&\;\;\;\;\; = \left[ n + \frac{1}{2} + \frac{1}{2} \left( \sqrt{
\mathstrut \eta + \textstyle \frac{1}{4} } - \sqrt{ \mathstrut
\eta } \right) \right] \pi. \;\;\;\;\; ( n = 0, 1, 2, \cdots )
\end{eqnarray}
If we solve this equation for $E$, the result is precisely
(\ref{exharm}); in this case, our semiclassical formula for energy
levels yields the {\it exact} results. Incidentally, for this
special case, we note that the exact results are reproduced even
if we use the WKB formula with the Langer modification, i.e.,
(\ref{langercon}).

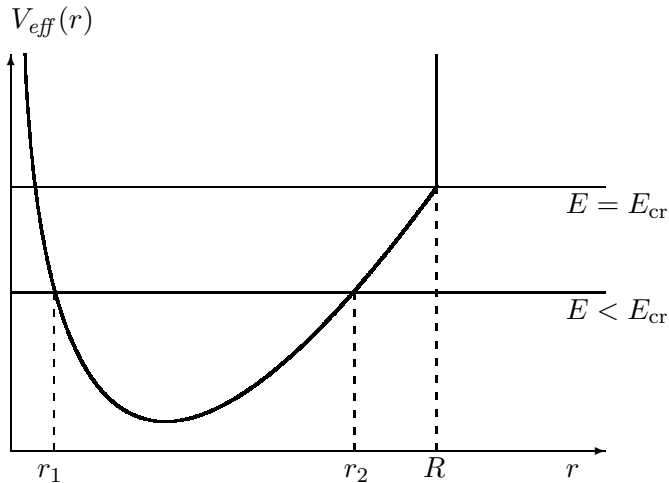
\begin{figure}[t]
\begin{picture}(230,180)
\put(210,0){$r$}
\put(210,100){$E = E_{\rm cr}$}
\put(210,60){$E < E_{\rm cr}$}
\put(0,170){$V_{\it eff}(r)$}
\put(10,0){$r_{1}$}
\put(126,0){$r_{2}$}
\put(156,0){$R$}
\put(0,10){\vector(0,1){150}}
\put(0,10){\vector(1,0){225}}
\put(0,110){\line(1,0){225}}
\put(0,70){\line(1,0){225}}
\put(161,10){\dashline[10]{3}(0,0)(0,100)}
\put(16.5,10){\dashline[10]{3}(0,0)(0,60)}
\put(130,10){\dashline[10]{3}(0,0)(0,60)}
\thicklines \bezier{5000}(5,160)(15,-90)(161,110)
\put(161,110){\line(0,1){50}}
\end{picture}
\caption{The shape of  $V_{\it eff}(r)$ for a circular billiard of
radius $R$. \label{fig3}}
\end{figure}

As a particularly simple example with nonzero interaction, let us
now consider a circular billiard potential (for general $B>0$),
i.e.,
\begin{equation} \label{bilv}
V(r) = \left\{
\begin{array}{cll}
0 &\;\;,& r<R \\
\infty &\;\;,& r>R.
\end{array}
\right.
\end{equation}
[This type of potential was previously considered in connection
with the virial coefficient calculation of the anyon gas
\cite{arovas2}]. The corresponding exact solution to (\ref{reom2})
is then the expression (\ref{exsol}), subject to the boundary
condition $R_{l}(r=R) = 0$, that is,
\begin{equation} \label{billiard}
F \left( a, b, z = \frac{eB}{4} R^{2} \right) = 0,
\end{equation}
when $a$ and $b$ here represent the same quantities as defined
above. Given information on the roots of (\ref{billiard}), one can
determine the exact energy eigenvalues. On the other hand, for the
semiclassical energy levels, we cannot use the formula
(\ref{WKBcon}) blindly --- for the effective potential as shown in
Fig. \ref{fig3}, the expression (\ref{WKB2}) is not appropriate.
That is, instead of (\ref{WKB2}), we must use the form
\begin{equation} \label{delta}
R_{l}^{\rm WKB}(r) \propto \frac{1}{\sqrt{p(r)}} \cos \left( -
\int_{r}^{r_{2}} dr' p(r') + \delta \right)
\end{equation}
with the phase $\delta \left( \neq \displaystyle \frac{\pi}{4}
\right)$ calculated to suit the present case. This in turn has the
consequence that we must add $\left( \delta - \displaystyle
\frac{\pi}{4} \right)$ to the right hand side of the quantization
condition (\ref{WKBcon}) (and of (\ref{langercon}) also). Here we
have two different situations depending on $E \geq E_{\rm cr}$ or
$E < E_{\rm cr}$, where $E_{\rm cr} = - \displaystyle
\frac{eB}{4\mu} (l-\alpha) + \frac{e^{2}B^{2}}{32\mu} R^{2} +
\frac{\eta}{2\mu R^{2}}$. For $E \geq E_{\rm cr}$, the second
turning point is at $r_{2}=R$ (i.e., at the infinite wall) and,
obviously, $\delta = \displaystyle \frac{\pi}{2}$ is the correct
choice. For $E < E_{\rm cr}$, however, we have $r_{2}<R$ (see Fig.
\ref{fig3}) and may consider another WKB wave function in the
classically forbidden region $r_{2}<r<R$,
\begin{equation} \label{mix}
R_{l}^{\rm WKB}(r) \propto \frac{1}{\sqrt{\kappa(r)}} \left\{ A
e^{ - \int_{r_{2}}^{r} dr' \kappa(r') } + B e^{ \int_{r_{2}}^{r}
dr' \kappa(r') } \right\}, \;\;\; \kappa(r) \equiv \sqrt{ 2\mu (
V_{\it eff}(r) - E ) }
\end{equation}
with the coefficients $A$ and $B$ chosen such that $R_{l}^{\rm
WKB}(r=R)=0$ may hold. Then the phase $\delta$ in (\ref{delta})
should be determined by using the (standard) WKB connection
formula with (\ref{mix}); this gives the value $\delta =
\displaystyle \frac{\pi}{4} + \cot^{-1} \left( 2 e^{ 2
\int_{r_{2}}^{R} dr' \kappa(r') } \right)$.

In the presence of the billiard potential (\ref{bilv}), we have
given above all the elements from which the exact and
semiclassical energy levels can be found, numerically at least. By
studying them, we have observed that the semiclassical
predictions, especially the ones based on our method (compared to
that based on the Langer modification of the potential), give
generically highly accurate results for a broad range of quantum
numbers. This should be evident from looking at Table \ref{tab1},
where the exact levels and the semiclassical predictions --- both
ours and that based on the Langer modification --- are compared
for some representative choice of parameters (for which we have
$r_{2}=R$ for all $n = 0, 1, 2, \cdots$). [With $R=5$ (but the
same values for the other parameters) we have $r_{2}<R$ for small
$n$; still, we have found the excellent overall agreement between
the semiclassical and exact energy levels].
\begin{table}[t]
\caption{Energy levels ($E_{n}$) for a circular billiard of radius
$R$ with nonzero $B$. The parameters were chosen as $l=2$, $\alpha
= \displaystyle \frac{1}{2}$, $\mu = \displaystyle \frac{1}{2}$,
$R=1$, $eB=2$. \label{tab1}}
\begin{center}
\tabcolsep = 1cm
\begin{tabular}{c|ccc}
\hline \hline n & Langer & Ours & Exact \\ \hline
0 & 18.4719 & 18.7389 & 18.7843 \\
1 & 57.9961 & 58.2515 & 58.2664 \\
2 & 117.2249 & 117.4775 & 117.4850 \\
3 & 196.1862 & 196.4378 & 196.4425 \\
4 & 294.8844 & 295.1355 & 295.1385 \\
5 & 413.3210 & 413.5717 & 413.5745 \\
6 & 551.4962 & 551.7468 & 551.7484 \\
7 & 709.4104 & 709.6608 & 709.6621 \\ \hline \hline
\end{tabular}
\end{center}
\end{table}

\section{s-wave semiclassical bound states}

In this section we specialize to the s-wave bound states of two
anyons. The relative wave function then depends on the coordinate
$r$ only, i.e., $\psi(\vec{r}) = \displaystyle \frac{1}{\sqrt{r}}
R_{0}(r)$, and from the Schr\"{o}dinger equation the function
$R_{0}(r)$ should satisfy the differential equation (see
(\ref{reom2}))
\begin{equation} \label{sreom}
\left\{ - \frac{1}{2\mu} \frac{d^{2}}{dr^{2}} + \frac{ \alpha^{2}
- \frac{1}{4} }{2\mu r^{2}} + \tilde{V}(r) \right\} R_{0}(r) = E
R_{0}(r),
\end{equation}
where $\tilde{V}(r) \equiv V(r) + \displaystyle \frac{eB}{4\mu}
\alpha + \frac{e^{2}B^{2}}{32\mu} r^{2}$ (the $l=0$ form of the
same function introduced in the previous section). For any
potential $\tilde{V}(r)$ which is regular at $r=0$, one may infer
the small-$r$ behavior of the solution $R_{0}(r)$ by studying the
solution of the simpler second-order differential equation,
obtained from (\ref{sreom}) with the replacement of the potential
$\tilde{V}(r)$ by the constant $\tilde{V}(0) = V(0) +
\displaystyle \frac{eB}{4\mu} \alpha$. In fact, it was an
analogous reasoning that picked the form (\ref{ex}) as a unique
allowed choice for $l \neq 0$. For s-wave states, on the other
hand, the form (\ref{ex}), i.e., $R_{0}^{\rm ex}(r) \propto
\sqrt{kr} J_{|\alpha|}(kr)$ does not represent the most general
small-$r$ behavior allowed for the exact solution of
(\ref{sreom}); here, one finds a {\it normalizable} wave function
even if $R_{0}(r)$ contains a piece corresponding to another
independent solutioin of the just-mentioned second-order
differential equation, that is, the form $\sqrt{kr}
N_{|\alpha|}(kr)$ ($N_{|\alpha|}$ is the Neumann function).
Therefore, for the small-$r$ behavior of the exact solution of
(\ref{sreom}), one may well consider the general form
\begin{equation} \label{sex}
R_{0}^{\rm ex}(r) \propto \sqrt{kr} \left\{ \cos\Theta \;
J_{|\alpha|}(kr) + \sin\Theta \; N_{|\alpha|}(kr) \right\},
\;\;\;\;\; \left( k = \sqrt{ 2\mu ( E - \tilde{V}(0) ) } \right)
\end{equation}
with an arbitrary $r$-independent angle $\Theta$.

In quantum mechanics with a singular potential, it is the theory
of self-adjoint extension of the related Hamiltonian that
determines what boundary conditions may be allowed. [A
particularly instructive example here is that of the
two-dimensional $\delta$-function potential problem, considered in
Ref. \cite{jackiw}]. From this theory follows the one-parameter
family of boundary conditions for our s-wave function
$\psi_{l=0}(r) = \displaystyle \frac{1}{\sqrt{r}} R_{0}(r)$
\cite{albeverio},
\begin{equation}
\lim_{r \rightarrow 0} \left\{ r^{|\alpha|} \psi_{l=0}(r) -
\tan\theta \left( \frac{2}{\rho} \right)^{2|\alpha|} \frac{
\Gamma(1+|\alpha|) }{ \Gamma(1-|\alpha|) }
\frac{d}{d(r^{2|\alpha|})} [ r^{|\alpha|} \psi_{l=0}(r) ] \right\}
= 0,
\end{equation}
where $\theta$, a dimensionless real number, is the self-adjoint
extension parameter, and $\rho$ just a (conveniently introduced)
reference scale. It is this general boundary condition that leads
to the small-$r$ behavior shown in (\ref{other}). One can now
translate the expression (\ref{other}) into the form (\ref{sex})
with the help of the identity $J_{-|\alpha|}(kr) =
J_{|\alpha|}(kr) \cos\pi|\alpha| - N_{|\alpha|}(kr)
\sin\pi|\alpha|$. Between the constant $\Theta$ in (\ref{sex}) and
the self-adjoint extension parameter $\theta$, we then find the
connection given in (\ref{cthetas}); this also tells us that
$\Theta$ depends on $k$ (and hence on energy $E$).

What would be the correct way of implementing the specific
small-$r$ behavior shown in (\ref{sex}) with semiclassical bound
states? Clearly, applying the standard WKB method with the
already-mentioned Langer modification trick would be unjustified
here. But we can still make use of the Friedrich-Trost approach
--- use the WKB wave function in the classically allowed region,
i.e.,
\begin{eqnarray} \label{sWKB}
&& R_{0}^{\rm WKB}(r) \propto \frac{1}{\sqrt{p(r)}} \cos \left\{
\int_{r_{1}}^{r} dr' p(r') + \phi \right\}, \\
&& \hspace{3cm} \left( p(r) = \sqrt{ 2\mu ( E - V_{\it eff}(r) )
}\; ; \;\; V_{\it eff}(r) = \tilde{V}(r) + \frac{ \alpha^{2} -
\frac{1}{4} }{2\mu r^{2}} \right) \nonumber
\end{eqnarray}
with the phase $\phi$ chosen such that this WKB wave function may
lead to an asymptotic form (for relatively large $kr$) in
agreement with that following now from the supposedly correct
expression in (\ref{sex}). The asymptotic expansion of (\ref{sex})
is easily found:
\begin{eqnarray} \label{saex}
R_{0}^{\rm ex}(r) &\sim& \left( 1 - \frac{\eta(\eta -
2)}{8(kr)^{2}} \right) \cos \left\{ kr - \left( |\alpha| +
\frac{1}{2} \right) \frac{\pi}{2} - \Theta
\right\} \nonumber \\
&& -\frac{\eta}{2kr} \sin \left\{ kr - \left( |\alpha| +
\frac{1}{2} \right) \frac{\pi}{2} - \Theta \right\} + O \left(
\frac{1}{(kr)^{3}} \right),
\end{eqnarray}
where $\eta \equiv \alpha^{2} - \displaystyle \frac{1}{4}$. As
regards the WKB wave function (\ref{sWKB}), a new problem arises
for $\eta<0$ (i.e., if $V_{\it eff}(r)$ contains an attractive
singular term at the origin) and it thus becomes necessary to
discuss the case with $|\alpha| < \displaystyle \frac{1}{2}$
separately from that with $|\alpha| \geq \displaystyle
\frac{1}{2}$. See below on this.

\begin{figure}[t]
\begin{picture}(230,180)
\put(210,50){$r$}
\put(210,100){$E$}
\put(0,170){$V_{\it eff}(r)$}
\put(151,50){$r_{2}$}
\put(0,10){\vector(0,1){150}}
\put(0,60){\vector(1,0){225}}
\put(0,110){\line(1,0){225}}
\put(156,60){\dashline[10]{3}(0,0)(0,50)}
\thicklines \bezier{5000}(5,10)(15,70)(100,90)
\bezier{5000}(100,90)(185,110)(200,160)
\end{picture}
\caption{The shape of $V_{\it eff}(r)$ with $\eta < 0$.
\label{fig2}}
\end{figure}
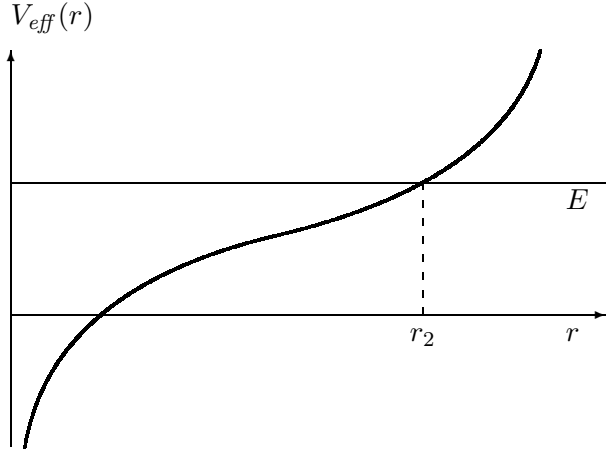

If $|\alpha|$ is lager than $\displaystyle \frac{1}{2}$ (i.e., for
$\eta>0$), the present s-wave WKB wave function will have
qualitatively the same structure as the non-s-wave WKB function
considered in (\ref{WKB}), and hence its asymptotic behavior by
the form (\ref{aWKB}) with the substitution $\eta = \alpha^{2} -
\displaystyle \frac{1}{4}$ now. Then, as we demand that it match
the asymptotic behavior in (\ref{saex}), the phase $\phi$ in the
WKB wave function (\ref{sWKB}) will be fixed to have the value
\begin{equation}
\phi = \frac{\pi}{2} \sqrt{ \alpha^{2} - \frac{1}{4} } - \left(
|\alpha| + \frac{1}{2} \right) \frac{\pi}{2} - \Theta, \;\;\;\;\;
\left( |\alpha| \geq \frac{1}{2} \right).
\end{equation}
[Note that, with $\Theta$ set to zero, this is nothing but the
formula (\ref{phi}). Also we have invoked the continuity of our
formula to include the case $|\alpha| = \displaystyle \frac{1}{2}$
here]. If $|\alpha| < \displaystyle \frac{1}{2}$ (i.e., for
$\eta<0$), on the other hand, we find the behavior $V_{\it eff}(r)
\rightarrow -\infty$ as $r \rightarrow 0$ and hence no classical
turning point near the origin. (See Fig. \ref{fig2}).
Thus, $r_{1}$ in the expression (\ref{sWKB}) is ambiguous. For
this case, one might be inclined to choose $r_{1}=0$; but, then,
the integral $\displaystyle \int_{0}^{r} dr' p(r')$ becomes
ill-defined. This difficulty is resolved if one chooses the point
$r_{1}$ to be not the origin but a certain (small) positive number
\cite{friedrich2}. The asymptotic behavior of the WKB wave
function (\ref{sWKB}) can then be described by the form
(\ref{aWKB}) again only if the constant $c$ there is now taken to
be
\begin{equation}
c = \sqrt{ |\eta| + (kr_{1})^{2} } - \sqrt{|\eta|} \ln \left(
\sqrt{ 1 + \frac{|\eta|}{(kr_{1})^{2}} } +
\frac{\sqrt{|\eta|}}{kr_{1}} \right), \;\;\;\;\; \left( {\rm for}
\; \eta = \alpha^{2} - \frac{1}{4} < 0 \right).
\end{equation}
Comparing such asymptotic expressioin with the behavior
(\ref{saex}), one immediately sees that the phase $\phi$ in this
case should be chosen
\begin{equation} \label{phi2}
\phi = c - \left( |\alpha| + \frac{1}{2} \right) \frac{\pi}{2} -
\Theta, \;\;\;\;\; \left( |\alpha| < \frac{1}{2} \right).
\end{equation}
The WKB wave function (\ref{sWKB}) for $|\alpha| < \displaystyle
\frac{1}{2}$, with this phase choice will have (in the region
beyond the point $r_{1}$) only a very mild dependence on the
cutoff $r_{1}$. Actually, we can do away with introducing this ad
hoc cutoff by changing our WKB wave function slightly. Let
$\bar{V}_{\it eff}(r)$ denote the small-$r$ approximation of
$V_{\it eff}(r)$, i.e.,
\begin{equation}
\bar{V}_{\it eff}(r) = \tilde{V}(0) - \frac{|\eta|}{2\mu r^{2}}.
\end{equation}
Then we would not sacrifice none of the desired properties of the
WKB wave function by changing the argument of cosine in
(\ref{sWKB}) to the expression
\begin{eqnarray}
&& \int_{0}^{r} dr' \left\{ \sqrt{ 2\mu ( E - V_{\it eff}(r') ) }
- \sqrt{ 2\mu ( E - \bar{V}_{\it eff}(r') ) } \right\} +
\int_{r_{1}}^{r} dr' \sqrt{ 2\mu ( E - \bar{V}_{\it eff}(r') ) } +
\phi \nonumber \\
&& \; = \int_{0}^{r} dr' \left\{ \sqrt{ 2\mu ( E - V_{\it eff}(r')
) } - \sqrt{ 2\mu ( E - \bar{V}_{\it eff}(r') ) } \right\} +
\sqrt{ |\eta| + (kr)^{2} } \nonumber \\
&& \;\;\;\;\; - \sqrt{|\eta|} \ln \left( \sqrt{ 1 +
\frac{|\eta|}{(kr)^{2}} } + \frac{\sqrt{|\eta|}}{kr} \right) -
\left( |\alpha| + \frac{1}{2} \right) \frac{\pi}{2} - \Theta,
\end{eqnarray}
where (\ref{phi2}) has been used. With $|\alpha| < \displaystyle
\frac{1}{2}$, this manifestly cutoff independent WKB wave function
shall be used in our discussions below.

In the presence of the second classical turning point $r=r_{2}$,
the WKB wave function found above can be used to obtain the s-wave
semiclassical energy levels. This part of argument is entirely
parallel to that in the previous section, and hence we may state
the result only. Assuming that the potential is smooth near the
second turning point and the WKB wave function in the region
$r>r_{2}$ contains a decreasing exponential only, the quantization
condition for energy levels reads, for $|\alpha| \geq
\displaystyle \frac{1}{2}$ (i.e., $\eta \geq 0$)
\begin{equation} \label{s-qc2}
\int_{r_{1}}^{r_{2}} dr \sqrt{ 2\mu ( E - V_{\it eff}(r) ) } =
\left[ n + \frac{1}{2} + \frac{1}{2} \left( |\alpha| - \sqrt{
\alpha^{2} - \textstyle \frac{1}{4} } \right) \right] \pi +
\Theta,
\end{equation}
and, for $|\alpha| < \displaystyle \frac{1}{2}$ (i.e., $-
\displaystyle \frac{1}{4} \leq \eta < 0$)
\begin{eqnarray} \label{s-qc}
&& \int_{0}^{r_{2}} dr \left\{ \sqrt{ 2\mu ( E - V_{\it eff}(r) )
} - \sqrt{ 2\mu ( E - \bar{V}_{\it eff}(r) ) } \right\} + \sqrt{
|\eta| + (kr_{2})^{2} } \hspace{2.5cm} \nonumber \\
&& \;\;\;\;\; - \sqrt{|\eta|} \ln \left( \sqrt{ 1 +
\frac{|\eta|}{(kr_{2})^{2}} } + \frac{\sqrt{|\eta|}}{kr_{2}}
\right) = \left( n + \frac{1}{2} + \frac{1}{2} |\alpha| \right)
\pi + \Theta.
\end{eqnarray}
Here it should be remarked that $\Theta$ itself is a function of
$E$, being related to the energy variable (and the self-adjoint
extension parameter $\theta$) by (\ref{cthetas}), It is through
the presence of this function $\Theta$ in (\ref{s-qc2}) and
(\ref{s-qc}) that the $\theta$-dependence enters the semiclassical
energy levels.

To check the accuracy of the above s-wave quantization formulas,
we may here consider the case $V(r)=0$, $B \neq 0$ (and with
$|\alpha| \geq \displaystyle \frac{1}{2}$ for simplicity). Then
the exact solution to (\ref{sreom}), which satisfies the
$\theta$-dependent boundary condition (\ref{other}) at the origin,
is
\begin{eqnarray} \label{thend}
&& R_{0}(r) = z^{ \frac{b}{2} - \frac{1}{4} } e^{-\frac{z}{2}}
F(a,b,z) \nonumber \\
&& \;\;\;\;\; + \tan\theta \left( \frac{2}{\rho}
\right)^{2|\alpha|} \left( \frac{eB}{4} \right)^{|\alpha|} \frac{
\Gamma(1+|\alpha|) }{ \Gamma(1-|\alpha|) } z^{ \frac{3}{4} -
\frac{b}{2} } e^{-\frac{z}{2}} F(1+a-b,2-b,z),
\end{eqnarray}
where $z \equiv \displaystyle \frac{eB}{4} r^{2}$, $a =
\displaystyle \frac{1}{2} ( 1 + |\alpha| + \alpha ) - \frac{2\mu
E}{eB}$, and $b = 1 + |\alpha|$. For this function to be
normalizable, we must further demand from the asymptotic
consideration that
\begin{equation}
\frac{\Gamma(b)}{\Gamma(a)} + \tan\theta \left( \frac{2}{\rho}
\right)^{2|\alpha|} \left( \frac{eB}{4} \right)^{|\alpha|} \frac{
\Gamma(1+|\alpha|) }{ \Gamma(1-|\alpha|) }
\frac{\Gamma(2-b)}{\Gamma(1+a-b)} = 0.
\end{equation}
Solving this equation for $E$ will give exact energy levels. We
may then compare the semiclassical energy levels obtained with the
help of (\ref{s-qc2}) against these exact values. [With $\theta
\neq 0$, the semiclassical formula does not produce exact energy
levels]. In Table \ref{tab2}, such comparison is made for some
representative choice of parameters. We find the high accuracy of
our semiclassical predictions, even for small $n$, very
impressive.
\begin{table}[t]
\caption{Energy levels ($E_{n}$) for the s-wave of the eigenstates
of $V=0$, but $B \neq 0$. The parameters were chosen as $\alpha =
\displaystyle \frac{2}{3}$, $\mu = \displaystyle \frac{1}{2}$, $eB
= 2$, $\theta = \displaystyle \frac{\pi}{4}$. \label{tab2}}
\begin{center}
\tabcolsep = 1.5cm
\begin{tabular}{c|cc}
\hline \hline n & Ours & Exact \\ \hline
-1 & 1.6667 & 1.6477 \\
0 & 3.3468 & 3.3440 \\
1 & 5.2337 & 5.2329 \\
2 & 7.1796 & 7.1793 \\
3 & 9.1480 & 9.1478 \\
4 & 11.1270 & 11.1269 \\
5 & 13.1120 & 13.1120 \\
6 & 15.1007 & 15.1007 \\
\hline \hline
\end{tabular}
\end{center}
\end{table}

\section{summary and discussions}

In this work we have developed the semiclassical theory of
two-anyon bound states, with careful consideration given on the
treatment of singular Aharonov-Bohm-type interactions between
anyons. The modified WKB method of Friedrich and Trost has proved
to be particularly effective for this problem. We have also
clarified the role of the self-adjoint extension parameter for
s-wave bound states within this semiclassical approach. For some
simple cases, we have been able to confirm that our semiclassical
formulas provide highly accurate energy levels over a broad
quantum number range. We expect this to be the case with more
general classes of interaction potentials. It should also be
possible to extend this semiclassical theory to the case of
two-anyon scattering states (in the absence of a background
magnetic field).

Results of the present semiclassical theory may be applied to
study certain features of the anyon gas. In the high-temperature
low-density limit, for instance, one usually resorts to the virial
expansion to study various thermodynamic properties. In the case
of the `free' anyon gas, the second virial coefficient has been
calculated in Refs. \cite{arovas2, moroz, kim3}; especially, in
Ref. \cite{kim3}, the effect of the self-adjoint extension
parameter on the virial coefficient has been considered also.
Based on the semiclassical understanding of two-anyon
bound/scattering states, one may extend this discussion to the
case of the anyon gas with some nontrivial 2-body potential, and
see what new features the presence of such 2-body interaction can
give rise to. Some of these issues are under investigation.

Also note that more general kinds of anyons, other than the ones
we discussed here, are possible. Anyons obeying so-called matrix
(or mutual) statistics \cite{wen, kim4} may prove to be relevant
in the discussion of multi-layered quantum Hall effect, and,
theoretically, particles obeying non-abelian statistics
\cite{wilczek2, bak} can also be contemplated. Bound states of
these kind of anyons may be discussed with the help of the
semiclassical theory analogous to the one considered in this
paper.

\section*{ACKNOWLEDGMENTS}

We would like to thank Seok Kim for interesting discussions. This
work was supported in part by the BK21 project of the Ministry of
Education, Korea, and the Korea Research Foundation Grant
2001-015-DP0085.

\end{document}